\documentclass{ws-procs975x65}

\begin{document}

\title{GAUGE THEORY OF GRAVITY WITH DE SITTER SYMMETRY\\ AS A SOLUTION TO THE COSMOLOGICAL CONSTANT\\ PROBLEM AND THE DARK ENERGY PUZZLE}

\author{PISIN CHEN$^{1,2}$}

\address{1. Department of Physics and Graduate Institute of Astrophysics \&\\
Leung Center for Cosmology and Particle Astrophysics\\ National Taiwan University, Taipei 10617, Taiwan, R.O.C.\\
$^*$E-mail: pisinchen@phys.ntu.edu.tw}

\address{2. Kavli Institute for Particle Astrophysics and Cosmology\\
SLAC National Accelerator Laboratory\\ Stanford University, Stanford, CA 94025, U.S.A.\\
E-mail: chen@slac.stanford.edu}

\begin{abstract}
We propose a solution to the longstanding cosmological constant (CC) problem which is based on the fusion of two existing concepts. The first is the suggestion that the proper description of classical gravitational effects is the gauge theory of gravity in which the connection instead of the metric acts as the dynamical variable. The resulting field equation does not then contain the CC term. This removes the connection between the CC and the quantum vacuum energy, and therefore addresses the {\it old} CC problem of why quantum vacuum energy does not gravitate. The CC-equivalent in this approach arises from the constant of integration when reducing the field equation to the Einstein equation. The second is the assumption that the universe obeys de Sitter symmetry, with the observed accelerating expansion as its manifestation. We combine these ideas and identify the constant of integration with the inverse-square of the radius of curvature of the de Sitter space. The origin of dark energy (DE) is therefore associated with the inherent spacetime geometry, with the smallness of DE protected by symmetry. This addresses the {\it new} CC problem, or the DE puzzle. This approach, however, faces major challenges from quantum considerations. These are the ghost problem associated with higher order gravity theories and the quantum instability of the de Sitter spacetime. We discuss their possible remedies. 

\end{abstract}


\bodymatter

\section{Introduction}\label{aba:sec1}
It is well-known that Einstein's general relativity (GR) allows for a cosmological constant (CC) which is a priori undetermined. The quantum vacuum energy resulting from the zero point fluctuations of the quantum fields, which is a natural consequence of the uncertainty principle, satisfies every aspect as a candidate for the CC. Yet numerically the quantum vacuum energy associated with the fluctuations at the Planck scale, $\sim 10^{112} {\rm eV}^4$, is about 124 orders of magnitude larger than the critical density of the universe. Evidently the quantum vacuum energy should not gravitate. Otherwise the universe would not have survived until now. This conflict between GR and quantum theory is the essence of the longstanding CC problem\cite{Weinberg}, which clearly requires a resolution. We shall refer to this as the {\it old} CC problem.

The dramatic discovery of the accelerating expansion of the universe in 1998\cite{Perlmutter,Riess} ushers in another chapter of the CC problem. The substance that is supposedly responsible for this accelerating expansion has been referred to as dark energy (DE). It has been customary to characterize DE in terms of its equation of state, $p=w\rho$, where $p$ is the pressure and $\rho$ the density of DE. According to GR, accelerating expansion can be accomplished if $w< -1/3$. Einstein's CC corresponds to $w=-1$. While the accuracy of the measurements to date still cannot resolve whether $w$ varies in time and whether it is exactly equal to $-1$, a constant DE and thus the CC remains the simplest and most likely answer. This, however, creates a new challenge for the CC problem. After finding a way, hopefully, to cancel the CC to 124 decimal points, how do we reinstate 1 to the last digit to make it nonzero but tiny? Let us call this the {\it new} CC problem, or the DE puzzle.

In this Letter, we propose a solution to the problems raised above by combining two existing ideas, that is, by invoking the gauge theory of gravity as a substitute of Einstein's GR as the foundation of gravity theory and by assuming de Sitter symmetry as the underlying group property of the universe. The gauge theory of gravity, which is quadratic in the curvature tensor, results in a field equation that is second order differential equation in the connection and therefore third order in the metric. The CC therefore does not appear in this field equation. This means the quantum vacuum energy is not a source of gravity. This proposal, however, does not remove the CC entirely. In order to reduce the field equation to the Einstein equation, one must perform a integration which then induces a constant of integration that behaves like a CC. To fix this constant, we assume that the universe satisfies the de Sitter symmetry. That is, we associate the constant of integration with the radius of curvature of the de Sitter space, a new fundamental length of the universe. We suggest that this new fundamental length plays a role in the action of the gauge theory of gravity.  

Neither of these two ingredients in our solution is new. Guided by the fiber bundle theory for gauge fields, Yang first formulated the gauge theory of gravity in 1974\cite{Yang1974}. This formulation was further investigated by various authors\cite{Thompson,Szczyrba,Gronwald}. In the aftermath of the discovery of accelerating expansion, recently Cook suggested the circumvention of the CC problem via the gauge theory of gravity\cite{Cook}. He observed that the quantum vacuum energy cannot be a source of gravity in the third order field equation derived from the gauge theory of gravity, while the DE responsible for the accelerating expansion of the universe can be identified with the constant of integration. The history of de Sitter symmetry dates further back. The notion that the physical laws are not invariant under the Poincare group but instead the de Sitter group was first proposed by Luigi Fantappie in 1954 and reinvestigated in 1968 by Bacry and Levy-Leblond\cite{Bacry}. In the post dark energy era, many authors have connected this notion with the observed accelerating expansion\cite{Guo,Aldrovandi,Cacciatori,Zee}. 

However, separately these two lines of solutions to the CC problem would be incomplete. While the gauge theory of gravity can indeed be devoid of the CC by construction, the CC reappears through the constant of integration. It may seem that one can at this point resort to the anthropic principle to settle its value. But as long as one still regards it as some form of stress energy-momentum, one would still be obliged to address the microscopic, or quantum, origin of this substance. Without a symmetry principle for the protection, this substance still has to be subject to quantum corrections, which may not help in preserving the desired smallness of the CC. The drawback of the solution that relies only on the de Sitter symmetry is more obvious. It simply does not address the old CC problem. That is, why doesn't the quantum vacuum energy gravitate? We will see that by fusing these two concepts together, the CC problem may be solved more properly.  

\section{Gauge Theory of Gravity and the Cosmological Constant}
In Yang's formulation of gravity, the affine connection $\Gamma^{\alpha}_{\mu\nu}$ on the fiber bundle is the dynamical variable which determines the curvature tensor
\begin{equation}
\label{curvaturetensor }
R^{\alpha}_{\beta\mu\nu}=\partial_{\mu}\Gamma^{\alpha}_{\beta\nu}-\partial_{\nu}\Gamma^{\alpha}_{\beta\mu}+\Gamma^{\alpha}_{\tau\mu}\Gamma^{\tau}_{\beta\nu}-\Gamma^{\alpha}_{\tau\nu}\Gamma^{\tau}_{\beta\mu}.
\end{equation}
For our purpose, we shall only consider the pure space. In close analogy with the Maxwell theory, the action of the gauge theory of gravity reads\cite{Cook}
\begin{equation}
\label{action}
   S_G=\kappa\int d^4x \sqrt{-g} \Big(R^{\alpha\beta\mu\nu}R_{\alpha\beta\mu\nu}+16\pi J_{\mu}^{\alpha\beta}\Gamma^{\mu}_{\alpha\beta}\Big),
\end{equation}
where $\kappa$ is the overall coefficient which will be determined later.
$J_{\mu}^{\alpha\beta}$ is the gravitational current defined as 
\begin{equation}
\label{current}
J_{\mu}^{\alpha\beta}=\frac{2G}{c^4}\Big[\nabla^{\alpha}\bar{T}_{\mu}^{\beta}-\nabla^{\beta}\bar{T}_{\mu}^{\alpha}\Big], 
\end{equation}
where $\nabla_{\alpha}$ is the covariant derivative and 
\begin{equation}
\bar{T}_{\mu}^{\alpha}= {T}_{\mu}^{\alpha}-\frac{1}{2}\delta_{\mu}^{\alpha}T
\end{equation}
is the Hilbert conjugate of $T_{\mu}^{\alpha}$ and $T=T_{\mu}^{\mu}$. Varying $S_G$ against $\Gamma^{\mu}_{\alpha\beta}$, we arrive at one of the field equation:
\begin{equation}
\label{fieldeq2}
\nabla_{\nu}R_{\alpha\beta}^{\mu\nu}=-4\pi J_{\alpha\beta}^{\mu}.
\end{equation}
This and the Bianchi identity, 
\begin{equation}
\label{fieldeq1}
\nabla_{\lambda}R_{\alpha\beta\mu\nu}+\nabla_{\nu}R_{\alpha\beta\lambda\mu}+\nabla_{\mu}R_{\alpha\beta\nu\lambda}=0,
\end{equation}
together determines the curvature tensor. The analogy of these equations to Maxwell's equations is again apparent. Note that Einstein's CC does not appear in (\ref{fieldeq1}). This fact can also be appreciated intuitively if we recall that the affine connection is related to the first derivatives of the metric,
\begin{equation}
\Gamma_{\alpha\beta\gamma}=\frac{1}{2}[\partial_{\gamma}g_{\alpha\beta}+\partial_{\beta}g_{\alpha\gamma}-\partial_{\alpha}g_{\beta\gamma}],
\end{equation}
then we see that the Bianchi identity is a third order differential equation in $g_{\mu\nu}$. This is in contrast with the Einstein equation which is second order in the metric. Since the covariant derivative of the metric is zero, a third order differential equation in $g_{\mu\nu}$ removes the CC term in principle. 

A pure space that is empty of energy-momentum satisfies the condition\cite{Yang1974} 
\begin{equation}
\nabla_{\gamma}R_{\alpha\beta}-\nabla_{\beta}R_{\alpha\gamma}=0,
\end{equation}
for the Ricci tensor $R_{\alpha\beta}$. This condition then reduces the gauge-Bianchi identity to 
\begin{equation}
\nabla_{\alpha}R^{\alpha}_{\beta\mu\nu}=0.
\end{equation}
Integrating this equation once over spacetime, we recover the Einstein equation with a constant of integration which has the same form as that of Einstein's CC. But in our case this constant should be associated with the boundary condition of the universe. 

\section{de Sitter Symmetry and the Dark Energy}
We now invoke the de Sitter symmetry and assume that the universe is inherently de Sitter, where the 4-spacetime is a hyperboloid in a 5-dimensional Minkowski space under the constraint
\begin{equation}
-x_0^2+x_1^2+x_2^2+x_3^2+x_4^2=l_{dS}^2,
\end{equation}
where $l_{dS}$ is the radius of curvature of the de Sitter space, or simply the de Sitter radius. The Hubble expansion of the universe is then viewed as a process that approaches the asymptotic limit of a pure space which is de Sitter in nature, evidenced by the current assessment that the CC-like DE substance has become dominant in the universe at late times: $\Omega_{DE}=\rho_{DE}/\rho_{cr}\simeq 0.75$, where the critical density $\rho_{cr}=3H_0^2/8\pi G=1.88\times 10^{-29}h^2 {\rm g/cm}^2$, with $h\equiv H_0/[100{\rm (km/s)/Mpc}]$ and $H_0$ is the Hubble constant. Based on this assumption the de Sitter radius is the asymptotic value of the Hubble distance, both in the gauge theory of gravity and in GR:
\begin{equation}
l_{dS}\simeq 1.33H_0^{-1}\sim 1.5\times 10^{28} {\rm cm}.
\end{equation}
Identifying the constant of integration as 3 times the inverse-square of the de Sitter radius, we then have
\begin{equation}
G_{\mu\nu}=-\frac{3}{l_{dS}^2}g_{\mu\nu},
\end{equation}
where $G_{\mu\nu}$ is the Einstein tensor. The only nontrivial component that satisfies this equation is a constant for the Ricci scalar,
\begin{equation}
\label{Ricci}
R=\frac{12}{l_{dS}^2}.
\end{equation} 
The local structure is then characterized by 
\begin{equation}
R_{\alpha\beta\mu\nu}=\frac{1}{12}[g_{\alpha\mu}g_{\beta\nu}-g_{\alpha\nu}g_{\beta\mu}]R,
\end{equation}
which confirms that the Kretschmann scalar is a constant in the de Sitter universe,
\begin{equation}
\label{Kretschmann}
R_{\alpha\beta\mu\nu}R^{\alpha\beta\mu\nu}=\frac{1}{6}R^2=\frac{24}{l_{dS}^4}.
\end{equation}
Based on this picture, the origin of DE is associated with the inherent spacetime geometry and not with vacuum energy or any other physical substance. Note that as an fundamental constant under de Sitter symmetry, $l_{dS}$ is not subject to quantum corrections. The smallness of DE is therefore protected by symmetry. 

In our definition of the gauge gravity action in (\ref{action}), we did not specify the overall coefficient $\kappa$. In general, the overall coefficient of a classical action does not affect the physics and is therefore irrelevant. It does matter, however, if we consider the quantum fluctuations around the classical state. We note that since the curvature tensor has dimension $[L]^{-2}$, any action term that is quadratic in the curvature tensor has its dimensionality entirely cancelled upon integration over the 4-spacetime. Thus the coefficient $\kappa$ must have the same dimension as the action itself. In comparison, since in the Einstein-Hilbert action the integration over the Ricci scalar has dimension $[L]^2$, its coefficient is $1/(16\pi G)$. Numerically, this factor is obtained by demanding that in the nonrelativistic limit GR reduces to the Newtonian gravity. But at the mean time since the gravity is weak, and therefore $1/G$ is large, quantum fluctuations of spacetime curvature in GR would be tiny at scales much larger than the Planck length. In the case of gauge theory of gravity, a natural choice for $\kappa$ would be the Planck constant $\hbar$, which has the right dimension. This, however, would imply that the the deviation of the curvature tensor from its classical minimum would be of the order unity at all scales, which is unacceptable. A different possibility is to introduce a length parameter $L$ so that $\kappa\propto L^2/G$. Following the same approach as in GR, we fix $\kappa$ numerically by demanding that the asymptotic de Sitter limit in the gauge theory of gravity is identical to that in GR. This can be done by comparing (\ref{Ricci}) and (\ref{Kretschmann}) above, which are the de Sitter solutions to the Einstein-Hilbert action and the gauge theory action, respectively. By demanding that the two theories have the same de Sitter limit, we find that $L$ is proportional to the de Sitter length $l_{dS}$, namely,
\begin{equation}
\kappa=\frac{l_{dS}^2}{96\pi G}.
\end{equation}

\section{Quantum Instability of de Sitter Spacetime}
One challenge of our approach is that de Sitter space may be inherently unstable. The issue of quantum instability of de Sitter space was investigated by various authors in the 1980s in the aftermath of the introduction of the concept of inflation. Abbott and Deser \cite{Abbott} have shown that de Sitter space is stable under a restricted class of classical gravitational perturbations. So any instability of de Sitter space may likely have a quantum origin. Ford \cite{Ford} demonstrated through the expectation value of the energy-momentum tensor for a system with a quantum field in a de Sitter background space that in general it contains a term that is proportional to the metric tensor and grows in time. As a result the curvature of the spacetime would decrease and de Sitter space tends to decay into the flat space. Similar conclusion has also been reached by Antoniadis, Iliopoupos, and Tomaras \cite{Antoniadis}.

We note that generically the expectation value of such energy-momentum tensor has the form
 \begin{equation}
\label{instability}
\langle T_{\mu\nu}\rangle\propto g_{\mu\nu}H^4(Ht),
\end{equation}
where $H$ is the Hubble parameter in the de Sitter metric, i.e.,
\begin{equation}
ds^2=dt^2-a^2(t)dx^2,
\end{equation}
with $a(t)=e^{Ht}$. 
According to (\ref{instability}), the decay time of this process is 
\begin{equation}
\tau\sim H^{-1}.
\end{equation}
In our case, this means the decay time is of the order of the de Sitter radius,
\begin{equation}
 \tau\sim l_{dS}\simeq 1.33H_0^{-1}. 
\end{equation}
Since the age of our universe is smaller than $l_{dS}$, we are still safe in observing the accelerating expansion in action. 

\section{Ghost Problem in High Order Gravity Theories}
Another major challenge of our approach is the ghost problem that is generally associated with higher order gravity theories such as ours. Ghost states are quantum states having negative norms. A quantum field theory is generally considered unacceptable if it contains ghost states, because negative norm implies negative probability. There are possible ways, however, to circumvent the ghost problem. 

Since our action is quadratic in $R_{\alpha\beta\mu\nu}$, which has mass dimension 2, the theory is conformally invariant. Analogous to the scalar field theory the conformal invariance can be spontaneously broken, which would induce a mass to the field. This means that our action will necessarily induce an Einstein term $R$, since it is proportional to $1/k^2$ in the propagator and therefore acts as a mass term. \cite{Kleinert} This implies that our theory goes to the Einstein GR at large distance. On the other hand, with $R_{\alpha\beta\mu\nu}\propto 1/k^4$, it is renormalizable at short distance. 
So in general the propagator of our gravity field becomes
\begin{equation}
G(k)=\frac{1}{k^4+ak^2+b}.
\end{equation}
It is clear that one of the poles of this propagator must be negative, and therefore there exists a ghost. 
However since such theory is scale invariant, we are free to choose the scale so that the ghost gets pushed to the Planck scale \cite{Kleinert}. Although this does not completely expel the ghost, the situation should become harmless. After all QED, for example, is also not free of divergence. The well-known Landau pole would appear when the energy goes to $m_e e^{1/\alpha}$. 

Another way to circumvent the ghost problem was recently proposed by Bender and Mannheim \cite{Bender}. They observed that the ghost problem may be originated from the fact that in the conventional quantum theory the Hamiltonian is assumed to be Dirac Hermitian, i.e., $H=H^{\dag}$. If the Hamiltonian is instead invariant under the more physical discrete symmetry of spacetime reflection, $H=H^{\cal PT}$, then the negative norm and probability would not occur to the propagator. Here parity ${\cal P}$ is a linear operator that performs space reflection and ${\cal T}$ is an antilinear operator that performs time reversal. If the energy spectrum of $H$ is real and positive, then the ${\cal PT}$ symmetry is unbroken and there exissts a reflection symmetry ${\cal C}$ which commutes with $H$, and also $[{\cal C}, {\cal PT}]=0$. Using this new inner product, it has been shown by Bender and others that the norm of a state is strictly positive. 

Before leaving this issue, we should like to mention that lately there has been a revival interest \cite{Wise} in the Lee-Wick theory \cite{Lee}, which also involves the ghost problem, albeit not so directly related to our issue. It should be fair to say that the ghost problem in higher order field theories, though serious, may not necessarily be incurable.    

\section{Discussion}
We see from the above discussion that the two assumptions for the solution of the CC problem, namely the gauge theory of gravity and the de Sitter symmetry, are knitted together due to the need to address both the old and the new CC problems. As often the case, a fundamental constant would manifest itself in various physical phenomena. Being an overall coefficient in the gauge gravity action, $l_{dS}$ does not appear in the field equations (\ref{fieldeq1}) and (\ref{fieldeq2}) and therefore would not affect the gravitational dynamics other than serving as the asymptotic limit of the cosmic expansion. It does, however, reveal itself in a curious way. Unlike GR in which the matter action is detached from the Einstein-Hilbert action, in gauge theory of gravity the matter field couples with the affine connection and is therefore an integral part of the total action, similar to the case of Maxwell theory. In this approach, Newton's constant acts as the gravitational charge. But if we insist on writing this matter-gravity coupling action separately from the curvature piece, then we have to absorb $\kappa$ into it and as a result we find that the actual matter-gravity coupling turns out to have the Newton's constant cancelled and is proportional to $l_{dS}^2$ only. What does this imply? On the one hand, the absence of Newton's constant in the matter action is the same situation as that in GR. On the other hand, unlike GR, the matter in the gauge theory of gravity couples to the geometry, i.e., the affine connection. However we now see that this coupling is mediated through the curvature of the de Sitter space. 

Another implication is that in our universe the Poincare symmetry should necessarily be replaced by the de Sitter symmetry. Such modification must be minute as $l_{dS}$ is astronomically large and therefore the deviation of spacetime from perfect flatness is small. Indeed there have been authors who look into the so-called de Sitter special relativity in recent years \cite{Guo}. The migration of spacetime symmetry from the Poincare group to the de Sitter group should in principle induce additional observable effects. These aspects are beyond the scope of this brief article and should be investigated separately. 

Our approach faces challenges based on quantum considerations. As we discussed, the problem of quantum instability of de Sitter space may not be fatal. Its decay time appears to be longer than the age of the universe. As for the problem of the ghost state, we argued that the negative pole may be pushed to the Planck scale and would therefore not be too harmful. These arguments are heuristic. To be sure, more in depth investigations are required before these problems can be resolved.

With regard to the gauge theory of gravity, in addition to the possibility of solving the CC problem, it may hopefully pave the way to the quantization of gravity and the unification with other gauge theories of interactions. As Yang himself commented in 1983: ``In \cite{Yang1974} I proposed that the gravitational equation should be changed to a third order differential equation. I believe today, even more than 1974, that this is a promising idea, because the third order equation is more natural than the second order one and because quantization of Einstein's theory leads to difficulties."\cite{Yang1983}.

\section*{Acknowledgments}
It is a pleasure to thank R. J. Adler, H. Kleinert, I. Antoniadis, G. t'Hooft, and my students Shu-Heng Shao and Nian-An Tung for helpful discussions. This research is supported by Taiwan National Science Council under Project No. NSC 97-2112-M-002-026-MY3 and by US Department of Energy under Contract No. DE-AC03-76SF00515.



\bibliographystyle{ws-procs975x65}
\bibliography{ws-pro-sample}

\begin{thebibliography}{9}
\bibitem{Weinberg} S. Weinberg, Rev. Mod. Phys. \textbf{},  (1989).
\bibitem{Perlmutter} S. Perlmutter et al., Astrophys. J. \textbf{517}, 565 (1999) [arXiv:astro-ph/9812133].
\bibitem{Riess} A. G. Riess et al., Astron. J. \textbf{116}, 1009 (1998) [arXiv:astro-ph/9805201].
\bibitem{Yang1974} C. N. Yang, Phys. Rev. Lett. \textbf{33}, 445 (1974).
\bibitem{Thompson} A. H. Thompson, Phys. Rev. Lett. \textbf{34}, 507 (1975).
\bibitem{Szczyrba} V. Szczyrba, Phys. Rev. D \textbf{36}, 351 (1987).
\bibitem{Gronwald} F. Gronwald and F. W. Hehl, arXiv: gr-qc/9602013.
\bibitem{Cook} R. J. Cook, arXiv:0810.4495v2 [gr-qc].
\bibitem{Bacry} H. Bacry, J-M. Levy-Leblond, J. Math Phys. \textbf{9}, 1605 (1968).
\bibitem{Guo} Han-Ying Guo, Chao-Guang Huang, Zhan Xu, Bin Zhou, Phys. Lett. A \textbf{331}, 1 (2004) [arXiv:hep-th/0403171].
\bibitem{Aldrovandi} R. Aldrovandi, J. G. Pereira, Found. Phys. \textbf{57}, 221 (2008) [arXiv:0711.2274]. 
\bibitem{Cacciatori} S. Cacciatori, V. Gorini, A. Kamenshchik, Ann. der Phys. \textbf{17}, 728 (2008) [arXiv:08073009].
\bibitem{Zee} A. Zee, lectures given at the Asia-Pacific Winter School on Cosmology, Taipei, Jan. 18-23, 2010.
\bibitem{Abbott} L. F. Abbott and S. Deser, Nucl. Phys. \textbf{B195}, 76 (1982).
\bibitem{Ford} L. H. Ford, Phys. Rev. D \textbf{31}, 710 (1985).
\bibitem{Antoniadis} I. Antoniadis, J. Iliopoupos, and T. N. Tomaras, Phys. Rev. Lett. \textbf{56}, 1319 (1986).
\bibitem{Kleinert} H. Kleinert, Phys. Lett. B \textbf{196}, 355 (1987); Private communications  (2010).
\bibitem{Bender} C. M. Bender and P. Mannhaim, J. Phys. A: Math. Theor. \textbf{41}, 304018 (2008); C. M. Bender and P. Mannhaim, Phys. Rev. Lett. \textbf{100}, 110402 (2008); P. Mannheim, arXiv:0912.2635 [hep-th].
\bibitem{Wise} B. Grinstein, D. O'Connell and M. B. Wise, Phys. Rev. D \textbf{77}, 025012 (2008).
\bibitem{Lee} T. D. Lee and G. C. Wick, Nucl. Phys. \textbf{B9}, 209 (1969); T. D. Lee and G. C. Wick, Phys. Rev. D \textbf{2}, 1033 (1970).
\bibitem{Yang1983} Chen Ning Yang, {\it Selected Papers, 1945-1980, With Commentary}, p.74 (W. H. Freeman, 1983). 
\end{thebibliography}

\end{document}